\documentclass[a4paper,UKenglish,cleveref, autoref, thm-restate,numberwithinsect]{lipics-v2021}

\hideLIPIcs  

\title{Calculational Proofs in ACL2s}

\author{Andrew T. Walter}{Khoury, Northeastern University, Boston, USA
  \and \url{https://www.atwalter.com/}}{walter.a@northeastern.edu}{https://orcid.org/0000-0002-7588-263X}{}

\author{Ankit Kumar}{Khoury, Northeastern University, Boston,
  USA \and \url{https://ankitku.github.io}}{kumar.anki@northeastern.edu}{https://orcid.org/0000-0001-9587-2861}{}

\author{Panagiotis Manolios}{Khoury, Northeastern University, Boston,
  USA \and \url{https://www.ccs.neu.edu/home/pete/index.html}}{p.manolios@northeastern.edu}{https://orcid.org/0000-0003-0519-9699}{}

\authorrunning{A. T. Walter and A. Kumar and P. Manolios} 

\Copyright{Andrew T. Walter and Ankit Kumar and Panagiotis Manolios} 



\category{} 

\relatedversion{} 

\acknowledgements{We sincerely thank Ken Baclawski, Raisa Bhuiyan,
  Harsh Chamarthi, Peter Dillinger, Robert Gold, Jason Hemann, Andrew
  Johnson, Alex Knauth, Michael Lin, Riccardo Pucella, Sanat Shajan,
  Olin Shivers, David Sprague, Atharva Shukla, Ravi Sundaram, Stavros
  Tripakis, Thomas Wahl, Josh Wallin, Kanming Xu and Michael Zappa for
  their help with developing and teaching with \hpc. We also thank all
  of the Teaching Assistants and students in CS2800, who are too
  numerous to list individually, who used and provided feedback on
  \hpc.}

\nolinenumbers 

\EventEditors{John Q. Open and Joan R. Access}
\EventNoEds{2}
\EventLongTitle{42nd Conference on Very Important Topics (CVIT 2016)}
\EventShortTitle{CVIT 2016}
\EventAcronym{CVIT}
\EventYear{2016}
\EventDate{December 24--27, 2016}
\EventLocation{Little Whinging, United Kingdom}
\EventLogo{}
\SeriesVolume{42}
\ArticleNo{23}

\usepackage{amsmath,amssymb,amsthm}
\usepackage{bbm}
\usepackage{paralist}
\usepackage[linesnumbered,ruled,vlined]{algorithm2e}
\usepackage{color,xcolor}
\usepackage{dsfont}
\usepackage[T1]{fontenc}
\usepackage[scaled]{beramono}
\usepackage{lstautogobble}
\usepackage{syntax}
\usepackage{dsfont,pifont}
\usepackage{float}

\newcommand{\p}[1]{\begin{normalfont}\frenchspacing\texttt{#1}\end{normalfont}}

\newcommand{\hpc}{CPC}
\newcommand{\ie}{\emph{i.e.}}
\newcommand{\eg}{\emph{e.g.}}
\newcommand{\chke}[1]{\varname{chk_e(#1)}}

\newcommand{\given}[1]{\varname{given(#1)}}
\newcommand{\cc}[1]{\varname{contract-completed(#1)}}
\newcommand{\export}[1]{\varname{export(#1)}}
\newcommand{\nil}{\varname{nil}}
\newcommand{\cts}{\mathds{C}}
\newcommand{\dts}{\mathds{D}}

\newcommand{\hints}{\mathds{H}}
\newcommand{\rel}{r}
\newcommand{\lhs}{\mathit{lhs}}
\newcommand{\rhs}{\mathit{rhs}}
\newcommand{\pr}{\mathit{proof}}
\newcommand{\proofbuilder}{proof-builder}
\newcommand{\rules}{\varname{Rules}}
\newcommand{\hyps}{\varname{Hyps}}
\newcommand{\uinst}{\varname{Uinst}}
\newcommand{\hyp}[1]{\varname{Hyp_{#1}}}
\newcommand{\conc}[1]{\varname{Conc_{#1}}}
\newcommand{\go}[1]{\varname{guards(#1)}}

\newcommand*{\funcfont}{\fontfamily{lmss}\selectfont}
\newcommand*{\codefont}{\ttfamily\small}

\newcommand{\hinteff}{\varname{hint-effect}}%
\newcommand{\xmark}{\ding{55}}%

\SetKwInput{KwInput}{Input}
\SetKwInput{KwOutput}{Output}
\SetKwFunction{ProveUsingHints}{ProveUsingHints}
\SetKwProg{Fn}{Function}{}{}
\SetKwFunction{CheckInductiveProof}{CheckInductiveProof}
\SetKwFunction{CheckNonInductiveProof}{CheckNonInductiveProof}
\SetKwFunction{CheckProofStep}{CheckProofStep}
\SetKwFunction{NonInductiveTranslate}{NonInductiveGenerate}
\SetKwFunction{InductiveTranslate}{InductiveGenerate}
\SetKwFunction{IndObsAndNames}{IndObsAndNames}
\SetKwFunction{DoProof}{DoProof}

\makeatletter
\DeclareRobustCommand{\varname}[1]{\begingroup\newmcodes@\mathit{#1}\endgroup}
\makeatother

\DeclareTextFontCommand{\funcfontify}{\funcfont}
\DeclareTextFontCommand{\codefontify}{\codefont}

\SetCommentSty{mycommfont}

\newcommand{\codify}[1]{\ensuremath{\mbox{\codefontify{#1}}}}

\lstdefinelanguage{handproofchecker-lang}
{
  morekeywords={Conjecture, Lemma, property, Prop},
  morecomment=[l]{;;},
}

\lstdefinestyle{proofStyle}{
language=handproofchecker-lang,
backgroundcolor=\color{white},
basicstyle=\scriptsize\ttfamily,
commentstyle=\ttfamily\itshape\color{gray},
tabsize=2,                    
showspaces=false,
showstringspaces=false,
autogobble=true
}

\graphicspath{ {./figs/} }

\begin{document}

\maketitle


\begin{abstract}
Teaching college students how to write rigorous proofs is a
  critical objective in courses that introduce formal reasoning. Over
  the course of several years, we have developed a
  mechanically-checkable style of calculational reasoning that we used
  to teach over a thousand freshman-level undergraduate students how
  to reason about computation in our ``Logic and Computation'' class
  at Northeastern University. We were inspired by Dijkstra, who
  advocated the use of calculational proofs, writing ``calculational
  proofs are almost always more effective than all informal
  alternatives, \ldots, the design of calculational proofs seems much
  more teachable than the elusive art of discovering an informal
  proof.'' Our calculational proof checker is integrated into ACL2s
  and is available as an Eclipse IDE plugin, via a Web interface, and
  as a stand-alone tool. It automatically checks proofs for
  correctness and provides useful feedback. We describe the
  architecture of the checker, its proof format, its underlying
  algorithms and provide examples using proofs from Dijkstra and from
  our undergraduate class. We also describe our experiences using the
  proof checker to teach undergraduates how to formally reason about
  computation.
\end{abstract}

\section{Introduction}

We present a calculational proof checker for ACL2s that was designed
to help undergraduate computer science students learn how to reason
about computation. In Spring of 2008, Manolios developed a course
entitled ``Logic and Computation,'' a required, freshman-level course
for computer science students whose goal was to introduce logic and
how it is used to reason about computation.  A major challenge in
teaching such a course is that students often cannot distinguish
between proofs and wishful thinking.  Therefore, we decided to use
ACL2s~\cite{dillinger-acl2-sedan,acl2s11} (an extension of
ACL2~\cite{acl2-web, acl2-car}), as it consists of a functional
programming language with formal semantics and an interactive theorem
prover.  The use of ACL2s allowed us to take advantage of the strong
intuition and extensive experience students have with programming to
discuss topics that include: programming language semantics, Boolean
algebra, writing specifications, equational reasoning, termination
analysis using measure functions, ordinal numbers, definitions,
algebraic data types, refinement, induction schemes, proof techniques
and formal proofs.

We used Dijkstra-inspired calculational proofs~\cite{dis90,ewds} from
the inception of the class and found them to be very effective
pedagogically. Our calculational proof format has gone through
continuous development, based on student feedback, our experiences
teaching the class, and experimentation with alternatives. For
example, an obvious approach that we tried (once) was to use the ACL2
proof builder, which allows one to take individual proof steps. Such a
capability is provided by almost all interactive theorem provers. We
discarded this idea because: (1) there was too much tool-specific
jargon, (2) students engaged in a lot of ``pointing,'' ``clicking''
and unstructured exploration, but not enough thinking, (3) students
found themselves with proof obligations that they could not easily
relate to the original conjecture, and (4) students did not learn
transferable skills. The calculational proof format we developed had
none of the above disadvantages and the following benefits: (1) it
includes a number of well-defined steps that are not ACL2s-specific
and help students structure proofs and avoid common pitfalls, (2) it
uses equational reasoning which leads to proofs that are easy to check
in a linear, local way, and (3) it is generally applicable to other
contexts requiring proofs.

After refining the proof format for a number of years, we developed a
proof checker, motivated by Manolios and Moore, who made the case for
mechanizing calculational proofs~\cite{man01}.  The main advantage of
mechanization is that students get immediate feedback and it
facilitates the automation of grading.  We deployed a Web version of
\hpc~(the Calculational Proof Checker) in Spring 2020, which accepts a
file containing function definitions, helper properties and
calculational proofs. It validates the proofs or identifies a list of
issues. It was later upgraded to check syntax in the browser itself,
so as to reduce load on the server. However, owing to its
server-client architecture, it required internet connectivity and
suffered from needlessly high response times, especially before
assignment submission deadlines. Hence, we recently developed an
Eclipse IDE plugin for writing and checking proofs locally. Finally,
for experts, there is a command-line version of \hpc\ that facilitates
scripting and supports enhanced debugging.  \hpc~is integrated with
the ACL2~Sedan (ACL2s) theorem
prover~\cite{dillinger-acl2-sedan,acl2s11}.  ACL2s is an extension of
the ACL2 theorem prover, whose authors were awarded the ACM Software
System Award in 2005~\cite{acl2-web, acl2-car, acl2-acs}.  ACL2s
extends ACL2 with an Eclipse IDE, a property-based modeling/analysis
framework, an advanced data definition framework
(\emph{Defdata})~\cite{defdata}, an automatic counterexample
generation framework
(\emph{cgen})~\cite{cgen,harsh-fmcad,disprovingTheorems}, and a
powerful termination/induction framework based on calling-context
graphs~\cite{ccg, ccg-cores, all-termination} and
ordinals~\cite{ManoliosVroon03, ManoliosVroon04, MV05}.  ACL2s
definitions and properties are ``typed'' and this helps students find
specification errors, as ACL2s performs a large number of analyses and
generates executable counterexamples highlighting errors.

Our contributions include: (1) Our \textbf{Calculational Proof
  Format}, which is designed for teaching, communicating, evaluating
and understanding proofs. It consists of a number of steps, such as
the identification and extraction of hypotheses. It supports
equational reasoning using a variety of relations. Our proof format
does not rely on keywords or syntax from any programming language,
making it applicable to a wide range of domains. Section~\ref{sec:ex}
provides a number of examples and an abstract grammar can be found in
Section~\ref{sec:format}.  (2) \textbf{\hpc}: Our proof checker is
based on ACL2s, which allows it to leverage the defdata
framework~\cite{defdata} to declare and reason about user defined data
types, the cgen framework for automatic counter-example generation for
invalid proof steps~\cite{cgen, harsh-fmcad, harsh-dissertation}, a
property based modeling/analysis framework and a vast library of
formalized results in the form of ACL2 books~\cite{books}. \hpc~proofs
can prove conjectures using ACL2s properties and vice-versa
recursively.  \hpc~comes with Eclipse IDE support, allowing for
locally writing and checking proofs.  The main algorithms of \hpc\ can
be found in Section~\ref{sec:algorithm} and the architecture of \hpc\
is described in Section~\ref{sec:architecture}.  It is also publicly
available.  (3) \textbf{Formalized and Validated Example Proofs}: We
have formalized, checked and made publicly available, several proofs
from Dijkstra's EWDs~\cite{ewds} and from the Logic and Computation class
at Northeastern University~\cite{repo}.

\section{Calculational Proof Examples with \hpc}
\label{sec:ex}
In this section, we show two calculational proof examples using
\hpc. The first example is from the Logic and Computation
class~\cite{lc} and the other is from EWD-1297~\cite{midfrac}.
Dijkstra's manuscripts are available online~\cite{ewds}.

\subsection{Sorting Homework Example}
A homework problem from our Logic and Computation class asks students
to prove the functional equivalence of quicksort and insertion sort.
This conjecture is written in ACL2s as follows.
\begin{lstlisting}[style=proofStyle,
  label=lst:isortqsortconjecture]
(property qsort=isort (x :tl)
  (== (qsort x) (isort x)))
\end{lstlisting}

\begin{figure}
\centering
\begin{minipage}{.45\textwidth}
\begin{lstlisting}[style=proofStyle,
  label=lst:isortqsortdefs,
  caption=Definitions and Helper Lemmas]
;; <<= is a total order on the ACL2s universe  
(definec <<= (x :all y :all) :bool
  (or (== x y) (<< x y)))
      
(definec insert (a :all x :tl) :tl
  (match x
    (() (list a))
    ((e . es) (if (<<= a e)
                  (cons a x)
                (cons e (insert a es))))))
                
(definec isort (x :tl) :tl
  (match x
    (() ())
    ((e . es) (insert e (isort es)))))
    
(definec less (a :all x :tl) :tl
  (match x
    (() ())
    ((e . es) (if (<< e a)
                  (cons e (less a es))
                (less a es)))))

(definec notless (a :all x :tl) :tl
  (match x
    (() ())
    ((e . es) (if (<<= a e)
                  (cons e (notless a es))
                (notless a es)))))

(definec qsort (x :tl) :tl
  (match x 
    (() ())
    ((e . es) (app (qsort (less e es))
                   (list e)
                   (qsort (notless e es))))))
                   
;; Useful abbreviations
(defabbrev A (x y z) (app x y z))
(defabbrev Q (x)     (qsort x))
(defabbrev I (x)     (isort x))
(defabbrev L (a x)   (less a x))
(defabbrev N (a x)   (notless a x))
(defabbrev F (x)     (first x))
(defabbrev R (x)     (rest x))
(defabbrev S (a x)   (insert a x))

;; Discovered lemmas  
(property isort-ordered (l :tl)
  (orderedp (isort l)))
        
(property isort-less (a :all l :tl)
  (== (isort (less a l))
      (less a (isort l))))

(property isort-notless (a :all l :tl)
  (== (isort (notless a l))
      (notless a (isort l))))
        
(property app-less-not-less (a :all l :tl)
  (=> (orderedp l)
      (== (append (less a l)
                  (cons a (notless a l)))
          (insert a l))))
\end{lstlisting}\hfill
\end{minipage}
\hspace{3em}
\begin{minipage}{.45\textwidth}
\begin{lstlisting}[style=proofStyle,
  label=lst:isortqsortproof,
  caption=Induction Step Proof]
;; Boolean manipulation to identify hypotheses
Exportation:
(=> (^ (tlp x)
       (!= x nil)
       (consp x)
       (=> (tlp (L (F x) (R x)))
           (== (Q (L (F x) (R x)))
               (I (L (F x) (R x)))))
       (=> (tlp (N (F x) (R x)))
           (== (Q (N (F x) (R x)))
               (I (N (F x) (R x))))))
    (== (Q x) (I x)))

;; Name all hypotheses
Context:
C1. (tlp x)
C2. (not (eq x nil))
C3. (consp x)
C4. (=> (tlp (L (F x) (R x)))
        (== (Q (L (F x) (R x)))
            (I (L (F x) (R x)))))
C5. (=> (tlp (N (F x) (R x)))
        (== (Q (N (F x) (R x)))
            (I (N (F x) (R x)))))

;; Use Modus Ponens to get concl of induction hyps
Derived Context:
D1. (tlp (R x)) { C2 }
D2. (tlp (L (F x) (R x))) { Def less }
D3. (tlp (N (F x) (R x))) { Def notless }
D4. (== (Q (L (F x) (R x)))
        (I (L (F x) (R x)))) { MP, C4, D2 }
D5. (== (Q (N (F x) (R x)))
        (I (N (F x) (R x)))) { MP, C5, D3 }
D6. (tlp (I (R x))) { D1, Def isort }
D7. (orderedp (I (R x)))
    { Lemma isort-ordered ((l (R x))), D6 }

Goal: (== (Q x) (I x))

Proof:
(Q x)
== { C2, Def qsort }
(A (Q (L (F x) (R x)))
   (list (F x))
   (Q (N (F x) (R x))))
== { D4, D5 }
(A (I (L (F x) (R x)))
   (list (F x))
   (I (N (F x) (R x))))
== { Lemma isort-less ((a (F x)) (l (R x))),
     Lemma isort-notless ((a (F x)) (l (R x))) }
(A (L (F x) (I (R x)))
   (list (F x))
   (N (F x) (I (R x))))
== { cons axioms, Def append, Def bin-app }
(A (L (F x) (I (R x)))
   (cons (F x) (N (F x) (I (R x)))))
== { Prop app-less-not-less
     ((l (I (R x))) (a (F x))), D6, D7 }
(S (F x) (I (R x)))
== { C2, Def isort }
(I x)
QED
\end{lstlisting}\hfill
\end{minipage}
\end{figure}

The ACL2s definitions of the sorting functions are shown in
Listing~\ref{lst:isortqsortdefs}.  Function definitions in ACL2s are
typed, \eg, \p{all}, \p{bool} and \p{tl} correspond to the ACL2s
universe, the Booleans and to the set of (true) lists, respectively.
The definitions should be self-explanatory and are accepted by ACL2s
without any other annotations or proofs.

Students are expected to prove \p{qsort=isort} with \hpc, using the
``professional method,'' a method taught in class that encourages the
use of custom notations and abbreviations, as shown in
Listing~\ref{lst:isortqsortdefs}.  The professional method allows one
to develop inductive proofs without committing to a particular
induction scheme; instead, an appropriate induction scheme is
discovered during the proof process, just like a pro.  Students are
encouraged to start with the most interesting part of a proof, which
in this case is the induction step. Students are then encouraged to
develop proofs in a top-down fashion so that they only prove lemmas
that actually turn out to be used in their main
proofs. Listing~\ref{lst:isortqsortdefs} includes the discovered
lemmas. 

Listing~\ref{lst:isortqsortproof} shows our \hpc\ proof of the
induction step, using the professional method. The proof is
constructed starting with \p{Proof} and trying to transform \p{(Q x)}
into \p{(I x)} by expanding the definition of \p{Q} (an abbreviation
for \p{qsort}), assuming that \p{x} is non-empty. Next, 
\p{Q}'s are converted to \p{I}'s by the magic of induction and then we
need a few lemmas to reason about properties of \p{I}. What remains is
to prove the discovered lemmas and to create a formal proof of the
induction step and the rest of the cases (which are not shown but are
in the supported materials). The induction scheme used is now obvious:
it is the induction scheme of \p{Q}, which is sound as \p{Q} was
automatically proven to terminate by ACL2s.  The stuff before the
\p{Proof} part is now rather routine, as students learn an algorithm
for generating induction schemes from function definitions that lead
to certain proof obligations (only the induction step is shown). The
\p{Exportation} step allows students to use only propositional logic
and simple type reasoning to transform the generated induction step
into a form that has as many hypotheses as possible and no
subexpressions of the form $\alpha \Rightarrow (\beta \Rightarrow
\gamma)$, as such expressions should be converted to $(\alpha \wedge
\beta) \Rightarrow \gamma$ (exportation). In the \p{Context}, each
hypothesis is given a name. Students are instructed to transform
induction hypotheses of the form $\alpha \Rightarrow \beta$ into
$\beta$ (so \p{C4}, \p{C5} get turned into \p{D4}, \p{D5}) in the
\p{Derived Context}, by establishing the hypotheses of each
implication and adding justification hints. They also can add any
other hypotheses they need in the \p{Proof} section. Steps in the
\p{Proof} section also require justifications, as shown.

Since \hpc\ is integrated with ACL2s, students use the counterexample
generation capabilities of ACL2s to automatically test their
discovered lemmas before proving them.  All properties in
Listing~\ref{lst:isortqsortdefs} were proved using \hpc, and their
proofs can be found in our repository~\cite{repo}.  Notice that in our
proof, we have applied lemmas \codify{isort-less} and
\codify{isort-notless}, with their corresponding instantiations,
simultaneously. Since both the applications are symmetric, and rewrite
different parts of the list, it helps to keep the proof short, yet
highlights the key steps of this proof. Proof steps can combine
several lemma applications in a single hint, as to keep the proof
short and human readable. We often also provide properties that are
proven in ACL2s directly and they can be used in the same way as
\hpc\ lemmas.

\subsection{EWD-1297}
\begin{figure}
  \includegraphics[scale=0.40]{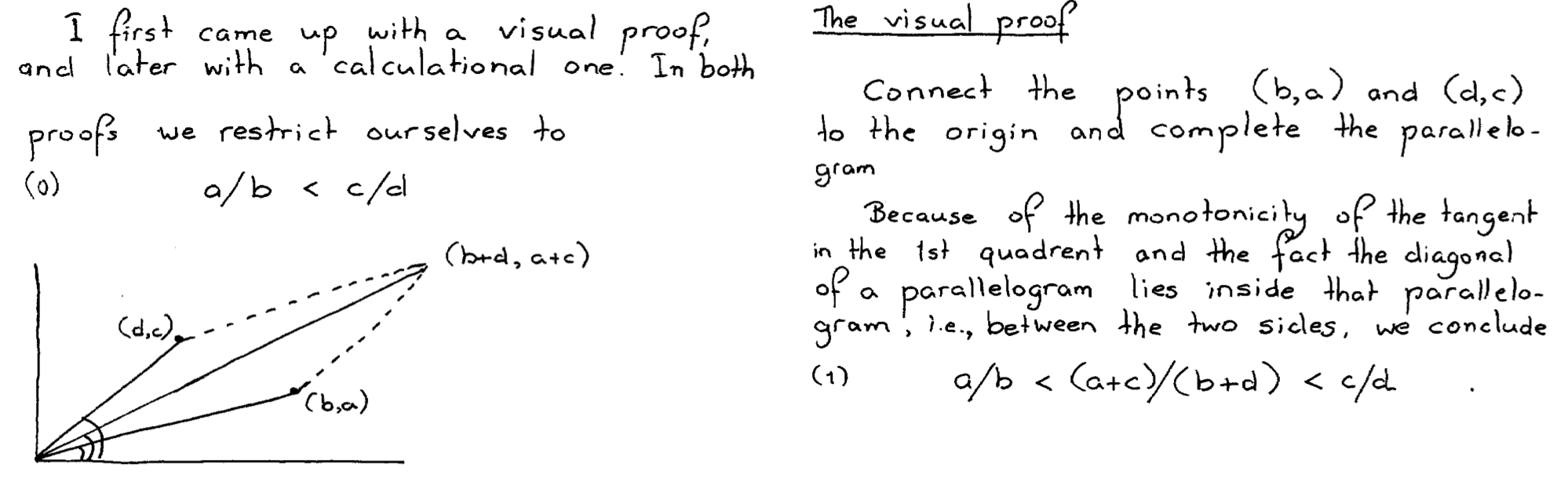}
    \vspace*{-3mm}
    \caption{Dijkstra's visual proof from EWD-1297}
    \label{fig:dewd1297-visual}
\end{figure}

Dijkstra describes a problem in EWD-1297~\cite{midfrac} which asks the
reader to construct a fraction whose value lies between two distinct
fractions $\frac{a}{b}$ and $\frac{c}{d}$. He restricts $a,b,c$ and
$d$ to be positive integers. He conjectured that $\frac{a+c}{b+d}$ is
a solution and gave a visual proof, as shown in
Figure~\ref{fig:dewd1297-visual}.
However, he points out that the visual proof is ``shaky'' and leaves
out several details. For example, it depends on the loosely stated
fact that the diagonals of a parallelogram lie inside it, \ie, between
two adjacent sides. But the proof of this ``fact'' might depend on the
demonstradum itself! He then gives a calculational proof, standing by
his famous quote: ``A picture may be worth a thousand words, a formula
is worth a thousand pictures.''~\cite{pics}
\begin{figure}
  \begin{subfigure}[t]{0.55\textwidth}
    \vskip 0pt
    \includegraphics[scale=0.26]{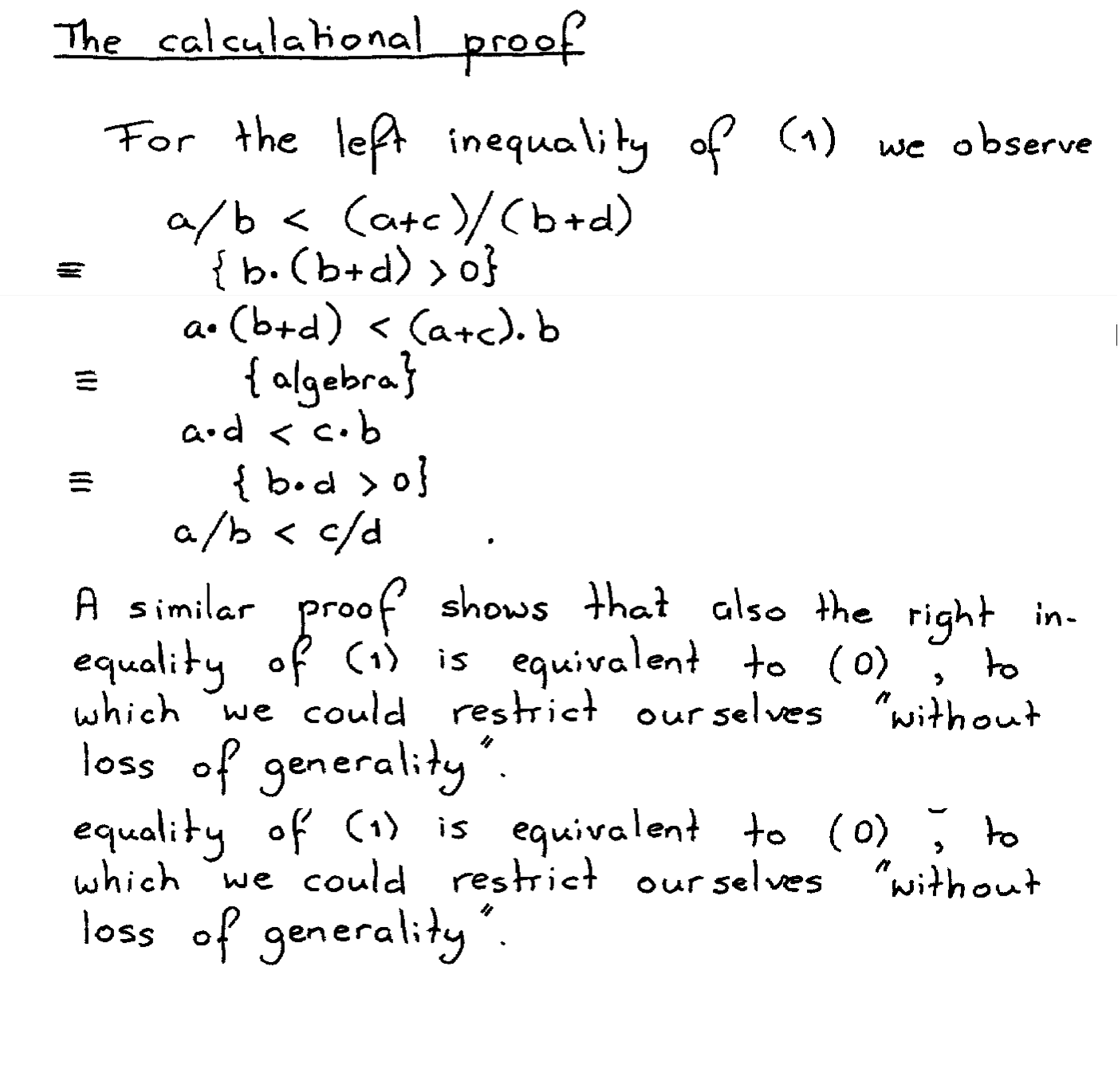}
    \vspace*{-6mm}
    \caption{Dijkstra's hand-written proof from EWD-1297}
    \label{fig:1297-1}

  \end{subfigure}
  \begin{subfigure}[t]{2\textwidth}
    \vskip 0pt
    \begin{lstlisting}[style=proofStyle]
Context:
C1. (posp a)
C2. (posp b)
C3. (posp c)
C4. (posp d)

Goal: (== (< (/ a b) (/ (+ a c) (+ b d)))
          (< (/ a b) (/ c d)))
Proof:
(< (/ a b) (/ (+ a c) (+ b d)))
== { Lemma multiply-<-fractions
     ((c (+ a c)) (d (+ b d))) }
(< (* a (+ b d)) (* (+ a c) b))
== { algebra }
(< (* a d) (* c b))
== { Lemma multiply-<-fractions }
(< (/ a b) (/ c d))
QED
\end{lstlisting}
\vspace*{-3mm}
\caption{Part of our formalization of\\ Dijkstra's proof from EWD-1297}
\label{lst:hpc-ewd1297}
  \end{subfigure}
  \caption{Side by side comparison of Dijkstra's proof and our proof
    formalization of EWD-1297~\cite{midfrac} }
  \vspace*{-6mm}
  \label{fig:ewd1297}
\end{figure}

In Figure~\ref{fig:ewd1297}, we show a side-by-side comparison of
Dijkstra's calculational proof and the corresponding part of our
formalization of his proof. One difference is that we use Lemma
\codify{multiply-<-fractions}, with appropriate instantiations,
instead of his hints, as our proof is formal. Our full formalization
appears in our repository~\cite{repo}, which includes the proof of
EWD-1297-2, the second part of the argument. Dijkstra did not prove
it, saying that the proof was similar, which is true.  Finally, we
proved the full conjecture EWD-1297 just by referring to the
previously proved conjectures.

Dijkstra was a fan of generalizing mathematical
results~\cite{gen,pythagoras,gen1} and often made the point that
calculational proofs, by clearly identifying assumptions, facilitate
such generalizations.~\footnote{Based on Manolios' personal
interactions with Dijkstra in classes and ATAC.} ACL2s can also be used to support such
generalization efforts. In Figure~\ref{fig:ewd1297}, we have four
context items positing that the variables $a,b,c$ and $d$ are positive
integers. As a first attempt to generalize this conjecture, we replace
each \codify{posp} by \codify{rationalp}, expanding the domain of the
variables to the rational numbers. \hpc\ produces the following
message:
\begin{lstlisting}[style=proofStyle]
--- Checking that completed statement passes contract checking... FAIL
Counterexample found when testing guard obligation:
(OR (NOT (AND (RATIONALP A) (RATIONALP B) (RATIONALP C) (RATIONALP D)))
    (NOT (EQUAL (+ B D) 0)))
(((D 0) (C 12/7) (B 0) (A -1/2))
 ((D 0) (C 1/3) (B 0) (A 122/875)) 
 ((D 0) (C 1) (B 0) (A 244/599)))
\end{lstlisting}
Obviously, \codify{b}, \codify{d} and \codify{(+ b d)} can not be $0$
since they are used as denominators in our goal. So, we add hypotheses
\codify{(!= 0 b)}, \codify{(!= 0 d)} and \codify{(!= 0 (+ b d))}. This
generalized conjecture has, as you would expect, the same proofs,
except for the extra hypotheses in the context and in proof
justifications. On running \hpc\ on our updated proof, we more
counterexamples and we keep going until we wind up with a
generalization, at which point we can use ACL2s again to removed
implied hypothesis. This process terminates with the generalization
shown in Listing~\ref{lst:ewd-1297-gen}, which \hpc\ proves. Note that the
only hypothesis we needed is the first hint of Dijkstra's proof!  The
full formalization of the generalized conjecture appears with
supplemental material and artifacts here~\cite{repo}.

\begin{figure}
\centering
\begin{minipage}{.47\textwidth}
\begin{lstlisting}[style=proofStyle,
  label=lst:ewd-1297-gen,
  caption=Generalized EWD-1297 proof]
Conjecture EWD-1297-5:
(=> (^ (rationalp a)
       (rationalp b)
       (rationalp c)
       (rationalp d)
       (< 0 (* b d)))
    (== (< (/ a b) (/ (+ a c) (+ b d)))
        (< (/ a b) (/ c d))))

Context:
C1. (rationalp a)
C2. (rationalp b)
C3. (rationalp c)
C4. (rationalp d)
C5. (< 0 (* b d))

Goal: (== (< (/ a b) (/ (+ a c) (+ b d)))
          (< (/ a b) (/ c d)))

Proof:
(< (/ a b) (/ (+ a c) (+ b d)))
== { C5, Lemma multiply-<-fractions
        ((c (+ a c)) (d (+ b d))), algebra }
(< (* a (+ b d)) (* (+ a c) b))
== { algebra }
(< (* a d) (* c b))
== { C5, Lemma multiply-<-fractions, algebra }
(< (/ a b) (/ c d))
QED

Conjecture EWD-1297-6:
(=> (^ (rationalp a)
       (rationalp b)
       (rationalp c)
       (rationalp d)
       (< 0 (* b d)))
    (== (< (/ (+ a c) (+ b d)) (/ c d))
        (< (/ a b) (/ c d))))

Context:
C1. (rationalp a)
C2. (rationalp b)
C3. (rationalp c)
C4. (rationalp d)
C5. (< 0 (* b d))

Goal: (== (< (/ (+ a c) (+ b d)) (/ c d))
          (< (/ a b) (/ c d)))

Proof:
(< (/ (+ a c) (+ b d)) (/ c d))
== { C5, Lemma multiply-<-fractions
        ((a (+ a c)) (b (+ b d))), algebra }
(< (* (+ a c) d) (* c (+ b d)))
== { algebra }
(< (* a d) (* c b))
== { C5, Lemma multiply-<-fractions, algebra }
(< (/ a b) (/ c d))
QED
\end{lstlisting}\hfill
\end{minipage}
\hspace{1em}
\begin{minipage}{.49\textwidth}
    \begin{lstlisting}[style=proofStyle,
    label=lst:ewd-1297-gen-full,
    caption=Complete proof of EWD-1297]
Conjecture EWD-1297-gen2:
(=> (^ (rationalp a)
       (rationalp b)
       (rationalp c)
       (rationalp d)
       (< 0 (* b d))
       (< (/ a b) (/ c d)))
    (^ (< (/ a b) (/ (+ a c) (+ b d)))
       (< (/ (+ a c) (+ b d)) (/ c d))))

Context:
C1. (rationalp a)
C2. (rationalp b)
C3. (rationalp c)
C4. (rationalp d)
C5. (< 0 (* b d))
C6. (< (/ a b) (/ c d))

Goal: (^ (< (/ a b) (/ (+ a c) (+ b d)))
         (< (/ (+ a c) (+ b d)) (/ c d)))

Proof:
(^ (< (/ a b) (/ (+ a c) (+ b d)))
   (< (/ (+ a c) (+ b d)) (/ c d)))
== { Conjecture EWD-1297-5,
     Conjecture EWD-1297-6, C5, C6, PL }
t
QED
\end{lstlisting}\hfill
\begin{lstlisting}[style=proofStyle,
  label=lst:ewd-1297-thm,
  caption=ACL2s proof of EWD-1297]
(property (a b c d :rational)
  :h (^ (< 0 (* b d))
        (< (/ a b) (/ c d)))
     (^ (< (/ a b) (/ (+ a c) (+ b d)))
        (< (/ (+ a c) (+ b d)) (/ c d)))
     :hints (("goal" :in-theory
                     (disable acl2::|(* (+ x y) z)|))))
\end{lstlisting}
\end{minipage}
\end{figure}

We note that when using ACL2s without providing hints, none of the
conjectures (original or generalized) in EWD-1297 are proved
automatically. However, as shown in Listing~\ref{lst:ewd-1297-thm},
there is an ACL2s proof of the generalized theorem that requires just
one hint (disabling a rule).  Figuring out what rule to disable can be a
lot of work, since there are dozens of applicable rules and many other
classes of hints that are potentially useful.  This is a common
problem when performing arithmetic reasoning inside ACL2s and other
interactive theorem provers that attempt to automate such reasoning
using rewriting and tactics because such methods are not complete (due
to undecidability) and users may have to intervene and guide the
theorem prover. The more complex the strategies employed, the more
expertise is required to effectively steer the proof.
The proof of EWD-1297 suggested some rewrite rules that we added to
ACL2s. It was easier to use \hpc\ to prove those theorems than using
ACL2s or the proof builder.  Our experience has been that
calculational proofs are easier to modify or repair, since they are
designed for human comprehension.  In contrast, trying to figure out
how to steer the theorem prover, was like playing whack-a-mole: the
ACL2 arithmetic libraries are robust, so efforts to steer the theorem
prover by providing explicit hints are often defeated by the remaining
dozens of rules conspiring to generate similar unproductive proof
attempts.

\section{Calculational Proof Format}
\label{sec:format}
Figure~\ref{fig:grm} shows an approximation of our calculational proof
grammar sufficient to understand the rest of this paper, especially
the main algorithms. \hpc\ uses Xtext, as described in
Section~\ref{sec:architecture} and the full grammar is in our
repository~\cite{repo}.
\begin{figure}[H]
  \centering
  \setlength{\grammarparsep}{4pt minus 5pt} 
  \setlength{\grammarindent}{9em} 
  \small
  \begin{grammar}
  <Proof> ::= <Type> $\mathcal{V}$ : $\mathcal{E}$ [Exportation: $\mathcal{E}$]
  [Contract Completion: $\mathcal{E}$] <Body> QED

  <Body> ::= <Simple> | <Inductive>

  <Simple> ::= [Context: <Ctx>] [Derived Context: <Dtx>] Goal: $\mathcal{E}$ Proof: <Seq>
  
  <Inductive> ::= Proof by: $\mathcal{E}$ [<ContractCase>] <BaseCase> [<InductionCase>]

  <ContractCase> ::= Contract Case $\mathds{N}$: <Simple> QED
  
  <BaseCase> ::= Base Case $\mathds{N}$: <Simple> QED
  
  <InductionCase> ::= Induction Case $\mathds{N}$: <Simple> QED
  
  <Type> ::= Conjecture | Property | Lemma | Theorem | Problem

  <Ctx> ::=  C$\mathds{N}$: $\mathcal{E}$ | <Ctx> C$\mathds{N}$: $\mathcal{E}$

  <Dtx> ::= D$\mathds{N}$: $\mathcal{E}$ | <Dtx> D$\mathds{N}$: $\mathcal{E}$

  <Seq> ::= $\mathcal{B}$ | <Seq> $\mathcal{R}$ \{<Hint>\} $\mathcal{B}$

  <Hint> ::= <Type> $\mathcal{V}$ [$\mathcal{S}$] | C$\mathds{N}$ | D$\mathds{N}$ |
  Def $\mathcal{F}$ | $\mathcal{A}$ | algebra | obvious | PL | MP | <Hint>,<Hint>
\end{grammar}
  \normalsize
  \caption{Grammar for our calculational proofs where, in the ACL2s
    universe, $\mathcal{V}$ is a fresh variable or natural number,
    $\mathcal{E}$ is an expression, $\mathds{N}$ is a natural number,
    $\mathcal{B}$ is a boolean expression, $\mathcal{R}$ is a binary
    relation on boolean expressions, $\mathcal{S}$ is an association
    list used to represent a valid substitution, $\mathcal{F}$ is a
    valid function name and $\mathcal{A}$ is an axiom. PL and MP stand
    for Propositional Logic and Modus Ponens hints
    respectively. Items in square brackets are optional.}
\label{fig:grm}
\end{figure}

We designed our proof format with a focus on teaching and
communicating proofs. Each Conjecture is named, so that it can be
referenced later. If Exportation and/or Contract Completion are
necessary, they need to be explicitly stated. This helps in finding
context items one might have missed when writing down the statement of
the Conjecture. The Context and Derived Context are listed separately
from the proof, which makes it easy to identify the exact context item
being used as a hint for a particular step. The proof body is linear,
\ie, a sequence of steps separated by hints that makes it easy to
identify any source of error due to wrong lemma application, missing
or incorrect hints. The grammar does not depend on keywords from any
programming language or proof assistant which makes it highly
transferable for use in other domains.

\section{Conjecture Checking and Proof Generation}
\label{sec:algorithm}
\hpc\ performs three ``phases'' of checking when it is run on a
document. The zeroth phase is a syntactic check to ensure that the
given proof follows the proof format described by our grammar, as
shown in Figure~\ref{fig:grm}. This check does not require making
calls to the underlying theorem prover. The first phase consists of
checks intended to identify semantic issues and produce useful output
for the user. After the syntax and semantics of a proof are validated,
the second phase checks its correctness by transforming each of the
given proofs into a sequence of instructions appropriate for use with
ACL2's \proofbuilder\ system. We then run each of these proofs through
ACL2s with the generated instructions and confirm that they pass.

A benefit of our multi-phase approach is the separation of concerns
enjoyed by the different phases of checking. The first two phases
interact with the user, focusing on giving actionable feedback, so as
to ensure that the given proof makes sense. The third phase interacts
with the underlying theorem prover for proving and generating
theorems. An example of why this is good, is that, in the absence of
contract completions, the behavior of the first phase does not affect
the argument for the correctness of \hpc, since the tool only reports
that a conjecture is valid if ACL2 can prove it. Before we describe
the three phases in detail, we discuss relevant background.

\noindent \textbf{ACL2 Theories}
A common way of controlling ACL2's behavior is by modifying the set of
\emph{rules} that it is allowed to use. There are various kinds of
rules, including general \codify{:rewrite} rules,
\codify{:type-prescription} rules that add type information that is
implied by the hypotheses that might be available at a particular
point in a proof, and \codify{:executable-counterpart} rules that
allow ACL2 to evaluate a function application when all of its
arguments have concrete values. ACL2 allows one to define and use sets
of rules, which it calls \emph{theories}. We make extensive use of
theories in \hpc, and describe the theories that we generate and use
as follows:
\begin{itemize}
\item \codify{arith-5} is the set of rules that are added after
  including the \codify{"arithmetic-5/top"} book in a vanilla ACL2
  instance. Put simply, this is a library developed by the ACL2
  community and distributed alongside ACL2 that contains rules for
  reasoning about arithmetic.
\item \codify{min} which consists of ACL2's minimal theory (which
  includes only rules about basic built-in functions like \codify{if}
  and \codify{cons}) plus \codify{(:executable-counterpart
    acl2::tau-system)}, \\ \codify{(:compound-recognizer
    booleanp-compound-recognizer)}, and \codify{(:definition
    not)}. The former of these three rules enables ACL2 to perform
  some type-based reasoning, and the latter two are often useful for
  reasoning about propositional logic.
\item \codify{arith} which consists of some basic facts about
  \codify{+} and \codify{*}.
\item \codify{type-prescription} which consists of any rules of
  type \codify{:type-prescription}
\item \codify{executable} which consists of any rules of type
  \codify{:executable-counterpart}.
\item \codify{contract} which is \codify{min} plus
  \codify{type-prescription} and any rules with names ending in
  \codify{"CONTRACT"} or \codify{"CONTRACT-TP"}.
\item \codify{min-executable} which is the union of the rules
  in \codify{min} and \codify{executable}.
\end{itemize}

The latter four theories are redefined every time \hpc\ checks a
conjecture, to ensure that any functions that may have been defined
since the last conjecture check are accounted for. Note that ACL2's
functions that operate on theories typically take in \emph{theory
  expressions}. For our purposes, a theory expression is either a list
of rule names, \codify{(theory '}$x$\codify{)} where $x$ is the name
of a previously defined theory or \codify{(union-theories }$x$
$y$\codify{)} which is a union of rules in theories $x$ and $y$.

\noindent \textbf{Guard Obligations and Type Checking} ACL2 is an
untyped language. However, it is possible to specify function
\emph{guards}, which are preconditions on the inputs to a
function. Violating function guards results in an error during
evaluation. Under certain circumstances ACL2 will attempt to prove
that the guards of a function will always hold in a particular context
(for example, when a function is called in the body of another
function with more restrictive guards). For the purpose of
explanation, ACL2s' type system hooks into the guard system in such a
way that a function defined using \codify{definec} (the standard way
to define functions in ACL2s) will have guards that assert that
provided inputs satisfy their specified types. The $\mathit{guard\
obligations}$ for an expression are the set of conditions that must be
true to satisfy the guards for every function call inside that
expression. We define a built-in function $\mathit{guards}$ such that,
$\go{e}$ is the conjunction of guard obligations for each function
call within an ACL2s expression $e$. Every ACL2s type has a
\emph{recognizer} function associated with it: a predicate of one
argument that returns \codify{true} when given a member of the
associated type, and \codify{false} otherwise. We refer to these
functions as \emph{type predicates} below.

\noindent \textbf{Contract Completion} We require that the statement a
user is trying to prove have an empty set of guard obligations
(equivalently, the guard obligations are all satisfied). To ensure
this in class exercises, we ask students to perform \emph{contract
completion} on given statements, before proving them.  Contract
completion refers to the process of determining the guard obligations
for a statement and adding appropriate hypotheses to satisfy
them. This of course changes the logical meaning of the proof
statement. From a pedagogical standpoint, we want to highlight the
correspondence between the statements we're proving and the code (the
executable bodies of the functions in the statement). By ensuring
contract completion, we eliminate all errors or counterexamples due to
guard violations.

Of course, it is always better if the original statement is already contract
completed. However, by doing a contract completion inside of \hpc\
first, we can confirm that it was done appropriately, \eg\ that only
the necessary hypotheses were added to satisfy the guard
obligations. After a contract completion is provided and validated, we
expect that the user will replace the original statement with the
contract completed one. We remind the user to do this by producing a
warning whenever they provide a \emph{non-trivial contract completion}
--- that is, a contract completed statement that is not syntactically
equivalent to the exported statement (if provided) or original
statement (if no exported statement was provided).

When the user provides a non-trivial contract completion for a
conjecture, we do not guarantee its correctness. Instead of generating
instructions to prove the conjecture inside of ACL2s, we tell ACL2s to
assume that the contract completed statement is true, which may cause
unsoundness issues if that is not the case. This design decision for
Phase I checks enables us to give more useful feedback regarding any
of the following proofs. For example, assume Conjecture A is followed
by Conjecture B and A has a nontrivial contract completion and it
passes Phase I checking. If we tell ACL2s to assume the contract
completed form of Conjecture A as valid, we can check Conjecture B and
see if it would pass our checks if the user replaced Conjecture A's
statement with its contract completed form.

\noindent \textbf{Hints}
Table~\ref{tab:hints} lists several types of hints for justifying
reasoning steps. Some hints have several aliases---for example,
\codify{algebra} and \codify{arith} are both aliases for
\codify{arithmetic}. Others are useful for improving readability of
the proof step. For example, \codify{MP} (\codify{Modus Ponens}) does
not add additional rules or hypotheses when checking a step or derived
context item, but we add it to indicate that a step or derived context
item is justified by the conclusion of an implication after satisfying
that implication's hypotheses. Each hint for a proof step gives rise
to zero or more of hypotheses ($\hyps$), $\rules$ and
$\varname{Lemma\ instantiations}$, used in proving the proof step. We
define functions $\mathrm{hyps}(h)$, $\mathrm{rules}(h)$ and
$\mathrm{instances}(h)$ to be the set of hypotheses, rules and lemma
instantiations (in a format amenable to ACL2) that a hint $h$ gives
rise to, respectively.  We also define function $\hinteff(h)$ which,
given a hint $h$, returns a triple
$(\mathrm{hyps}(h),\mathrm{rules}(h),\mathrm{instances}(h))$ as described
in Table~\ref{tab:hints}.

\begin{table}[]
\begin{tabular}{c|c|c|c|l}
  Hint & Hypotheses  & Rules & Lemmas & Comments \\
  \hline
  $\codify{C}i$ & \checkmark & \xmark  & \xmark  &  Context items\\
  \codify{D}$i$ & \checkmark & \xmark  & \xmark  &  Derived Context items\\
  \codify{def }$\mathit{foo}$ & \xmark & \checkmark  & \xmark  & Enables rules regarding the Definition of $\mathit{foo}$ \\
  \codify{cons axioms} & \xmark & \checkmark  & \xmark  &  Enables Lemmas related to lists\\
  \codify{arithmetic}  & \xmark & \checkmark  & \xmark  & Enables Lemmas
                                                 related to arithmetic \\
  \codify{evaluation}  & \xmark & \checkmark  & \xmark  &  \\
  \codify{lemma }$\mathit{foo}$  & \xmark & \xmark  & \checkmark & A substitution may be provided
  \\
  \codify{MP }  & \xmark & \xmark  & \xmark & improves readability \\
  \codify{PL }  & \xmark & \xmark  & \xmark & improves readability
\end{tabular}
\caption{A table showing what is returned by each hint.}
\label{tab:hints}
\end{table}

\noindent \textbf{Phase 0}:
Phase 0 consists entirely of syntactic checks. As we will discuss in
more detail in the Architecture section~\ref{sec:architecture}, we use
the Xtext~\cite{xtext} system to generate a parser from our proof
format grammar (described in the previous section). Xtext
automatically generates many syntactic checks and provides
error-handling capabilities, but we extend these checks to produce
specialized error messages in certain situations.

\noindent \textbf{Phase I}: We describe Phase I checks using algorithms
that generate helpful error messages, given semantically incorrect
proofs. We use the $\textit{chk}_e$ function to denote checking if a
Boolean ACL2s expression is true, otherwise returning an appropriate
error. $\varname{guards}$ returns the guard obligations of its
argument. $\varname{type-predicatep}$ tells whether an expression is
an ACL2s function application term where the function is a type
recognizer. $\mathit{given}$ checks if its argument has been provided
in the proof. $\mathit{export}$ exports an ACL2s
expression. $\varname{contract-completed}$ checks if the given ACL2s
expression's contracts appear in its hypothesis.
$\vdash_{\mathit{th}}s$ checks whether $s$ is can be proved by ACL2s in
theory $\mathit{th}$. Each of the functions described above is defined
in ACL2s. Phase I is intended to identify errors with enough
specificity to allow us to generate clear, actionable and useful error
messages.

\begin{algorithm}
  \small
  \caption{Semantic checks for a single proof step or derived context item}
    \label{alg:proof-step}
  \Fn{\CheckProofStep{$\cts$, $\hints$, $\lhs$, $\rhs$, $\rel$}}{
    \KwIn{$\cts$ is a set of context items, $\hints$ is a set of
      hints, $\lhs,\rhs \in \mathcal{E}$ are both ACL2 expressions, and
      $\rel \in \mathcal{R}$ is an ACL2 relation.}
    $\hyps,\rules,\uinst \gets \hinteff(\hints)$\;
    $R \gets \{ c \ |\ c \in \cts \wedge \varname{type-predicatep}(c)\}$ \;
    $G \gets \go{ \hyps \wedge R\ \Rightarrow \lhs\ \rel\ \rhs}$ \;
    $\chke{\vdash_{\codify{min+contract+}\rules\codify{+}\uinst} G\ \wedge
      \rules\ \wedge \hyps\ \wedge R\ \Rightarrow \lhs\, \rel\, \rhs}$
    \tcp*[l]{Proving a step with calculated guard obligations relieves
      the user from the need to provide type hints}
    $\chke{\vdash_{\codify{full}} \wedge \cts \Rightarrow \wedge G}$\; }
        \normalsize
\end{algorithm}

Algorithm~\ref{alg:proof-step} can be considered to be the heart of
\hpc. It checks whether $\lhs$ is related to $\rhs$ using relation
$\rel$ in a given context $\cts$ and with hints $\hints$. This
algorithm is used to check the validity of each derived context item
as well as each proof step. Any context items specifying type
information (variable $R$ in the algorithm) are automatically assumed,
so the user need not provide type justification hints. We tell
students that once they have performed contract checking for the
conjecture, they no longer obligated to justify types for terms
occurring in function contracts, lemma instantiations or implications,
which mimics Dijkstra's style of proofs. Notice that we subsequently
determine the guard obligations of the step (after adding type
recognizer applications), add the guard obligations as an additional
hypothesis when proving the step holds and afterwards ask ACL2s to
prove that they hold given all of the context and derived context
items. This helps us identify cases where the user does something
incorrect from a ``types'' perspective, while also freeing the user
from having to justify types, enabling them to focus on the key ideas
of the proof.
\begin{algorithm}[]
        \small
  \label{alg:semantics-simple-phase1}
  \caption{Semantic checks on a simple proof.}
    \label{alg:semantics-simple-phase1}
  \Fn{\CheckNonInductiveProof{$\pr$}}{ \KwIn{$\pr$ is a non-inductive
      (simple) proof. $\cts$ is a set of context items. $\dts$ is a
      set of pairs of derived context items and corresponding set of
      hints. $S$ is the proof statement, whose exportation (if given)
      is $E$. $C$ (if given) is the contract completion of $E$ (if
      given) otherwise $S$. $G$ is the goal.
      $P \in \langle Seq \rangle$ is the sequence of proof steps separated by
      relations (relating proof steps) and hints (justifying each
      step). $(\lhs,\rel,H,\rhs) \in P$ is a single proof step in $P$. }
    
    \tcc{Exportation Checks}
    \eIf
    {$\given{E}$ \tcp*[h]{E is given?}}  {
      $\chke{\vdash_{\codify{min}}E \equiv S}$ \tcp*[l]{E is proved equivalent
        to S in min-theory?}  }{ $E \gets S$\; }
    $\chke{\export{E} \equiv E}$ \tcp*[l]{E is fully exported?} 
    \tcc{Contract Completion Checks}
    \eIf{$\given{C}$ \tcp*[h]{C is given?}}{
    $\hyp{E},\conc{E} \gets E$ \tcp*{Split E into a set of hypotheses and a conclusion}
    $\hyp{C},\conc{C} \gets C$ \tcp*{Split C into a set of hypotheses and a conclusion}    
    $\chke{\conc{C} \equiv \conc{E}}$ \;
    $\chke{\hyp{C} \subseteq \hyp{E}}$
    $\chke{\vdash_{\codify{min}} \wedge \hyp{C} \equiv (\hyp{E} \wedge \go{E}) }$ 
    \tcp*[l]{Contract completion hyps equivalent to guard
      obligation and hyps of E?}
  }{
    $C \gets E$\;
    $\hyp{C},\conc{C} \gets C$ \tcp*[h]{Split C into a set of hypotheses and a conclusion}    
  }
  $\chke{\cc{C}}$ \tcp*[l]{C is contract completed?}
  \If{$\given{G}$}{
    $\chke{\vdash_{\codify{min}}\conc{C} \equiv G}$ \tcp*[l]{Conclusion is proved equivalent to Goal?}
  }
  \tcc{Context Checks}
  $\chke{\vdash_{\codify{min}} (\wedge \cts \Rightarrow G ) \equiv C }$ \;
  \tcc{Derived Context Checks}
  \ForEach{$(d,H) \in \dts$} {
     \tcc{Each Derived Context item is checked to be equivalent to true}
    $\CheckProofStep{$\cts$,H,d,\codify{true},\codify{equal}}$ \;
    $\cts \gets \cts \cup \{ d \}$ \;
  }
  \If{$(\nil,\_) \in \dts \vee G \in \dts$}{
    \Return \tcp*[l]{Exit early because hypotheses are UNSAT or they imply G}
  }
  $\chke{\given{G}}$ \tcp*[l]{otherwise G must be provided}

  \tcc{Proof Step Checks}
  \ForEach{$(\lhs,\rel,H,\rhs) \in P$} {
    $\CheckProofStep{$\cts$,H,$\lhs$,$\rhs$,$\rel$}$\;
  }
  $\chke{\vdash_{\codify{min}}\bigwedge\limits_{(\lhs,\rel,H,\rhs)\in P}\lhs \Rightarrow G}$ \tcp*{Final check that Proof steps imply the Goal}
  }
        \normalsize
\end{algorithm}
Algorithm~\ref{alg:semantics-simple-phase1} checks a non-inductive
(simple) proof. For the purpose of presentation, we are showing only
the core set of checks it performs. The implementation does additional
boilerplate checking to make sure that the types of expressions we are
dealing with are correct, as well as certain additional checks to
identify potential issues.  For example in our implementation, unless
the user derives \codify{nil}, we check whether the conjunction of the
context items is SAT, since an unsatisfiable set of hypotheses likely
indicates a mistake when it's not used to trivially prove the proof
statement.

For inductive proofs, it is important to make sure that given an
induction term, the proof obligations fulfilled by the user correspond
exactly to the proof obligations generated internally in ACL2. We
check this by searching for a bijection between the two sets of proof
obligations with respect to propositional equality. Once a bijection
is shown, we check each of the proofs using
Algorithm~\ref{alg:semantics-simple-phase1}.

\noindent \textbf{Phase II}:
If a proof passes through Phase I without generating any errors, we
attempt to either generate instructions that tell ACL2 how to prove
the conjecture, or if it has a non-trivial contract completion, we
tell ACL2 to assume that the conjecture holds. We'll focus here on the
former case, since the latter is only intended to provide best-effort
feedback and we make no claims about correctness if a conjecture has a
non-trivial contract completion anyways. Before we can discuss
generating instructions, we need to present some background about the
ACL2 \proofbuilder.

\noindent \textbf{ACL2 \proofbuilder}: ACL2 is designed to be highly
automatic; when it fails to prove a theorem, the user typically proves
helpful lemmas or provides hints that modify or override its
behavior. This mode of operation is not ideal in the context of proof
checking, since we already have a detailed outline for a proof, and we
want ACL2 to stick to that outline. For this reason, we use ACL2's
\proofbuilder\ functionality, which allows us to command the theorem
prover's behavior at a much lower level. The \proofbuilder\ operates
in a manner similar to an interactive proof assistant like
Coq~\cite{coqart} or Isabelle~\cite{isabelle-hol}: there is a
\emph{proof state} consisting of a stack of goals, each of which
contains a set of hypotheses and a statement to be proved, and one
provides \emph{instructions} that operate on the goal stack. These
instructions range in fineness, with coarse instructions like
\codify{prove} (attempt to prove the current goal entirely
automatically with ACL2's full power) to fine instructions like
\codify{dive} (look at a subexpression in the current statement to be
proved). ACL2's documentation provides information about many of the
available \proofbuilder\
instructions~\cite{xdoc-proof-builder-commands}.  For \hpc\ we
developed several new \proofbuilder\ instructions, many of which are
variants of existing instructions that succeed where the existing
instructions would fail. For example, \codify{:retain-or-skip} is
exactly like the built-in \codify{:retain} instructions, except that
it will succeed even when all of the existing hypotheses are retained
(producing no change in the \proofbuilder\ state). Many instructions
have similar behavior that is desirable when a human is interacting
directly with the \proofbuilder, but that is not when automatically
generating instructions.

\noindent \textbf{Generating Instructions}: To check a calculational
proof, we generate a sequence of \proofbuilder\ instructions based on
the given proof. We ask ACL2s to prove the proof statement using this
sequence of instructions. This is useful in two ways. First, the
soundness of \hpc\ is reduced to the soundness of ACL2's
\proofbuilder\ system. Second, this means that in principle, one can
perform a proof in \hpc\ that might be challenging to do in ACL2s, by
taking the generated proof form and using it in ACL2s. Recall the
structure of a calculational proof: we have a statement, an
exportation (if needed), context (if needed), derived context items
(if generated) and a sequence of steps separated by relations and
justifications. The order in which \proofbuilder\ instructions are
generated mirrors this structure.

The generated \proofbuilder\ instructions are executed in ACL2 which
will either produce an error or run smoothly, thereby turning the
proof conjecture into an ACL2s property. In the latter case, the
generated \proofbuilder\ instructions act as a witness to the validity
of the conjecture.

\section{System Architecture}
\label{sec:architecture}
Our \hpc\ system consists of three parts: (1) a user interface which
allows the user to write a proof file, (2) a parser that performs some
basic well-formedness checks on a given proof file and (3) a proof
checking backend that validates a syntactically well formed proof for
correctness.

\begin{figure}[t]
  \includegraphics[width=\textwidth]{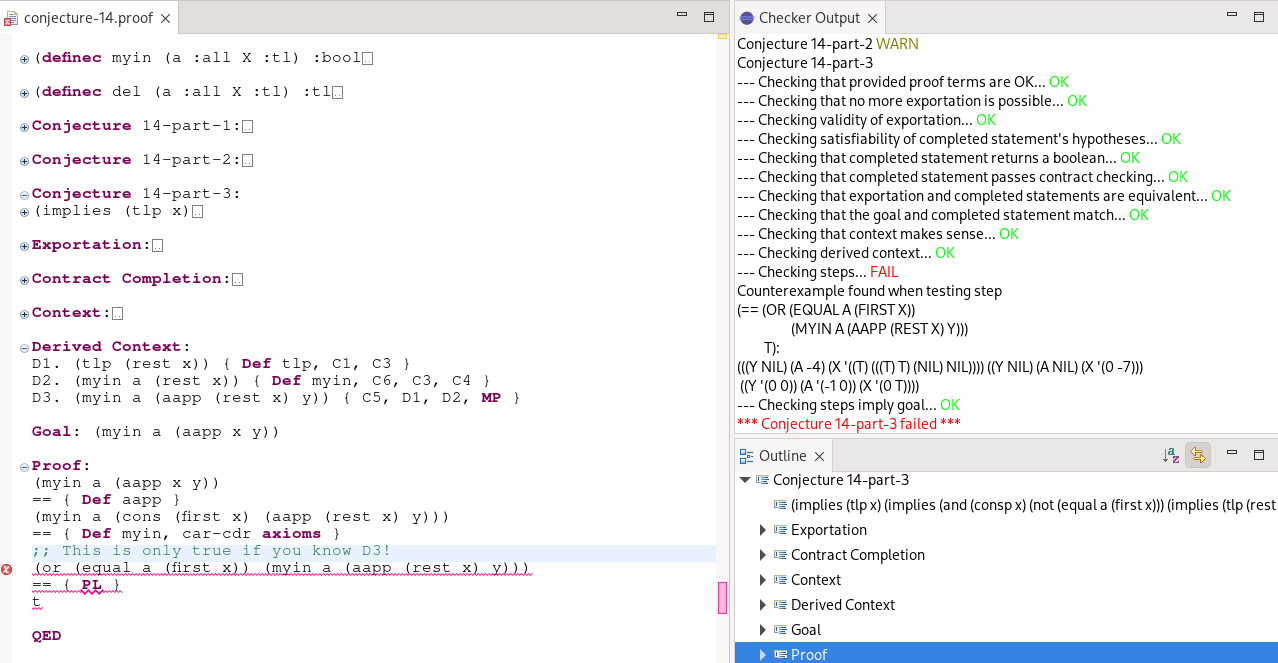}
  \caption{A screenshot of our Eclipse plugin showing an
    incorrect version of our example ``conjecture-14'' proof. Note
    that the counterexample listed here is not a counterexample to the
    correctness of the step, but rather to the statement that the step
    is true due to the provided justification.}
  \label{fig:eclipse-plugin-screenshot}
\end{figure}

We provide two primary interfaces for \hpc: a Web-based editor and an
Eclipse plugin. The Web-based editor requires no setup on the user's
part but has fewer features and is slower than the Eclipse plugin,
since much of its functionality comes from making requests to a remote
server. The Eclipse plugin is faster and has more features, but
requires more effort to set up locally.
Figure~\ref{fig:eclipse-plugin-screenshot} shows a screenshot of our
Eclipse plugin. We discuss features of the user interface in more detail.

In both the interfaces, the results of checking a proof are
communicated to the user in two ways: (1) by producing a simplified
list of the checks that is performed for each proof and their status,
with some additional details if the check failed, as is shown in the
``Checker Output'' panel on the top-right-hand side of
Figure~\ref{fig:eclipse-plugin-screenshot}, and (2) by marking
particular areas of the user's conjectures with
\textit{diagnostics}. An error diagnostic can be seen in
Figure~\ref{fig:eclipse-plugin-screenshot}, visualized by the red
underlining of part of the conjecture shown on the left-hand side of
the screenshot. Certain checks may give rise to lower-severity
diagnostics, like warnings or informational notes.

We spent significant time and effort in developing user-facing
messages that are easy to understand, useful and actionable. In fact,
this is the raison d’etre for having Phase I. But we also owe much to
the years of usage that \hpc\ has accumulated and the experience we
have gained perfecting it.

The Eclipse plugin is equipped with several features to improve the
proof-editing experience. Parts of a proof, or even entire proofs, can
be folded (hidden) when desired to maximize the screen space available
for content of interest (using $\oplus$ and $\ominus$ buttons to the left of text
seen in the screenshot) which makes it easy to browse a large proof
file. An outline view is available that allows one to easily identify
and navigate to specific areas of a document. Certain keywords like
Conjecture, Proof, Def etc... are highlighted. Highlighting of matched
parentheses and basic code completion are provided. The Eclipse plugin
also provides links to online documentation describing the conjecture
format we use, and answers to frequently asked questions.

We use the Xtext system~\cite{xtext} to implement both the parser and
a portion of both the user interfaces. Xtext is a Java-based framework
for developing tooling for domain-specific languages. By providing a
grammar, Xtext generates not only a parser but also IDE features like
code completion, syntax highlighting and code folding. The default
behavior that Xtext provides can be modified by providing Java
code. We configure Xtext to generate both an Eclipse plugin and
supporting functionality for the CodeMirror Web-based editor. Phase 0
checks described in the previous section are implemented by the parser
that Xtext generates automatically from the grammar we provide, though
we augment its error-handling abilities to produce more helpful error
messages in certain conditions. Xtext provides the ability to define
arbitrary checks (called ``validators'') for a document written in a
DSL. These validators take in an abstract syntax tree (AST), process
it and mark different parts of the AST with various diagnostics as
needed. Xtext handles mapping these errors back to the text of the
document and rendering them to the user. To integrate our proof
checker backend with Xtext, we define a validator that encodes the
conjecture document in a form readable by our backend, runs the
backend in a subprocess, captures and parses the output from the
backend, and translates that output into Xtext diagnostics.

The backend is responsible for checking the correctness all of the
proofs contained in a document. It performs Phase I and Phase II
checks as described in the previous section and uses the ACL2s Systems
Programming methodology that we have
developed~\cite{walter-acl2-systems-programming}. That is, most of the
backend's code is written in Common Lisp and runs in a Common Lisp
environment that has ACL2 loaded. It interacts with ACL2 through a set
of interface functions, making it easier to identify which code
interacts with the theorem prover and reducing the risk of
accidentally modifying ACL2 internals, thereby making ACL2
unsound. This potential unsoundness is mitigated for our purposes as
\hpc\ produces ACL2s proofs that can be checked by running them
through an unmodified ACL2s instance.

The backend takes in a syntactically correct
proof document that consists of ACL2 expressions interleaved with
conjectures. We define data structures in Common Lisp that track
metadata attached to expressions or parts of a conjecture, which
informs us where certain diagnostics should appear. The proof document
is represented as a sequence of these data structures. In brief, when
given a proof document, the backend performs some setup, sequentially
evaluates each element of the proof document while collecting
diagnostics, and then reports the diagnostics in an XML document. This
XML data is captured and consumed by our Xtext validator and is
translated into Xtext diagnostics, which are visible to the user
through whichever frontend they are using.

\section{Our Experience Teaching With \hpc}
We have several years of experience using \hpc\ to teach students how
to write calculational proofs. The following features have proved to
be very useful. (1) \textbf{Mechanization:} students get immediate
feedback, allowing them to quickly learn how to distinguish proofs
from wishful thinking. (2) \textbf{Syntax Checking:} Students get
instantaneous syntax-related feedback. Due to this, the TAs almost
never have to deal with incorrect syntax in student submissions.  (3)
\textbf{Useful Feedback:} \hpc\ produces helpful feedback on proof
steps to help students correct any mistakes in their proof, \eg,
incorrect lemma instantiations, missing hints etc.  We continuously
improve the error messages and validation due to student feedback. (4)
\textbf{ACL2s Integration:} This allows us to provide lemmas for
certain assignments and allows students to test any lemmas and
specifications they write using the $\mathit{cgen}$ framework before
actually investing any effort in formal proofs.

The use of \hpc\ also improved the educational experience for
students, as the immediate feedback \hpc\ provides allows them to
transition more quickly from basic proof questions to proof strategy
considerations. For a small subset of students, who previously may not
have realized they needed help, \hpc\ helps them realize they need
help without having to wait for a grading cycle to get that feedback.

Despite these advantages, we did get feedback that the online version
of \hpc\ was sometimes too slow (taking several minutes) when checking
particularly long proofs. This can be attributed to the many
validation queries required to check proofs and overloaded
servers. Another request was to enable proof checking even when
students were not online.  Both of these requests are addressed by the
Eclipse plug in.

\section{Related Work}
\label{sec:rel}
Calculational proofs were popularized by Dijkstra~\cite{dis90},
Gasteren~\cite{vg90} and Gries~\cite{gri91} in the early 90s. This
style of proof was not new at the time. Dijkstra credits
Hoare~\cite{dream} and Feijen~\cite{feijen} for introducing him to
calculational proofs. Calculational proofs have since become popular
among computer scientists for their readability and mathematical
rigor. Lamport~\cite{lamport1995write} advocated structured proofs
using hierarchical numbering and proof constructs from TLA+. Robinson
and Staples proposed window inference~\cite{rswindow} as a more
flexible alternative to calculational proofs, as it allows selection
and transformation of subexpressions, while preserving
equivalence. Grundy described window inference in terms of a sequent
formulation of natural deduction and generalized it to allow for
general transformational proofs~\cite{grundy96trx}. Following this,
Back, Grundy and Wright proposed structured calculational
proofs~\cite{back97,grundy96} as a way to incorporate the nested
structure of natural deduction proofs~\cite{gen35,pra65} into
calculational proofs, thereby improving on both readability and
browsability.

At the turn of the twenty first century, Manolios and Moore argued for
mechanized checking of calculational proofs~\cite{man01}. Noting that
(1) Calculational proofs are amenable to mechanical
verification~\cite{mverif} and (2) proofs often follow from a
syntactic analysis of the formal statement of the
demonstradum~\cite{demo}, they challenged the calculational proof
community to build a program to check proofs in a sound manner. Today,
there exist several such programs. We will discuss them according to
the following criteria :\\
\textbf{Social aspects} refer to the ease of understanding,
communicating and teaching proofs. A calculational proof should
neither be bloated with, nor be restricted by programming language
specific syntax. It should focus solely on communicating the key ideas
(of a proof) as clearly as possible. Dijkstra favored having clear and
explicit justifications about the context which (1) conveys key proof
ideas and (2) avoids repetition of
subexpressions~\cite{notation}. Separating context from proof helps in
this regard, since context items can be listed explicitly and used in
justifications. We show later in the paper, how context items help in
coming up with generalized conjectures.\\
\textbf{Formal aspects} refer to the level of checking done by proof
checkers. Typically, proof checkers lie on a spectrum, ranging from no
checking whatsoever, to full checking of the validity of proofs,
usually with the help of an underlying theorem prover.\\
\textbf{Utilitarian aspects} refer to the utility of a proof checker
as a proof assistant. We believe that a proof checker should not only
be able to verify proofs, but also interactively help users with the
construction of proofs.\\
The Mizar~\cite{mizar92} (Multi-Sorted with Equality) system,
developed in the early 90s by Andrzej Trybulec and his group in
Bialystok, is an early effort to mechanize checking calculational
proofs. The Mizar language is formal, yet highly readable due to its
declarative nature and a rich syntax which allows for close
resemblance of proofs to mathematical vernacular. However the system
is not very interactive. It depends on light weight
automated-reasoning and single rule application to justify proof
steps. John Harrison added a ``Mizar Mode''~\cite{harrison-mizar} on
top of the HOL theorem prover providing a version of Mizar language
which is more interactive as well as a corresponding proof checker
which is more powerful. A similar approach was taken to implement Isar
for Isabelle~\cite{isar,mizarisar} and Mizar Light for HOL
Light~\cite{wiedijk2001mizar} to produce more interactive and formal
systems. Leino and Polikarpova~\cite{lepo13} extended Dafny (a SMT
solver based auto-active program verifier) with program-oriented
calculations (poC) to write structured calculational proofs. Program
derivation is a programming technique that derives programs from
specifications by means of formula manipulation, which is essentially
a calculational proof. Tesson et al.~\cite{tesson11} developed a set
of lightweight tactics to support program derivation in Coq.  Vazou et
al.~\cite{vazou2018theorem} use Liquid Haskell, a program verifier for
Haskell which uses refinement types, to calculate efficient
programs. However, all of these systems lack in social aspects in the
following ways (1) Context is not made explicit in the proof (2)
Relations allowed between proof steps are limited (Mizar and the
program derivation tools only support equalities while Mizar Mode for
HOL, Isar and Dafny have a fixed set of transitive relations).

On the other side of the spectrum of formal aspects lies
Mathedit~\cite{back07mathedit}, which allows writing structured calculational
proofs, with the ability to expand and contract subderivations. However,
derivation steps are checked only for syntactic and semantic requirements on
the derivation, which is not a formal verification of the correctness of the
proof itself.

There exist several notable tools that check proofs in a limited theory. Mendes
and Ferreira~\cite{mendes18} presented a proof of concept system that verifies
hand-written (from a pen-based input device) calculational proofs in
propositional logic, using Isabelle/HOL as the backend prover to check proof
steps. Kahl~\cite{kahl18} introduced the CalcCheck system to input, verify and
grade calculational proofs from the book ``A Logical Approach to Discrete
Math"~\cite{ladm93}. Carl et al. presented Diproche~\cite{diproche}, an
automated proof checker for proofs of elementary mathematical exercises written
in a controlled fragment of German, the ``Diproche language”. Barwise's
Hyperproof system~\cite{barwise1994hyperproof} checked proofs combining
graphical and sentential information using a fixed set of logical rules.

Apart from proof checking, there is a large body of work in proof
visualization, including but not limited to Grundy's
ProofViews~\cite{grundy97} package for browsing a proof interactively
along its hierarchical structure, and the Omega
system~\cite{siekmann1999lomegaui} to visualize proofs interactively
using proof trees. However, we believe that proof visualizers have
limited applicability above and beyond pedagogy, whereas our work is
more applicable. To the best of our knowledge, \hpc~is the best
calculational proof checker in every aspect we discussed.

We end with previous work related to ACL2s, which is an extension of
ACL2. The ACL2s core libraries and installation instructions are part
of the ACL2 distribution~\cite{acl2s-web}. The Gitlab
group~\cite{acl2s-group} contains a collection of projects related to
ACL2s, including the scripts project which includes scripts for
standalone ACL2s implementations~\cite{acl2s-scripts}.  ACL2 and ACL2s
have been used in industry to formalize, analyze and reason about a
variety of complex industrial systems~\cite{hunt2017industrial}. ACL2s
examples include: (1) the formalization and automated analysis of
system requirements for safety-critical avionics systems, leading to
the ASSERT tool that has been used on numerous industrial
projects~\cite{requirements-2018, assert-2017}, (2) the validation of
conformance of hardware devices to standardized networking protocols
using hardware-in-the-loop fuzzing, which introduced the idea of
enumerative types with constraints, allowing us to solve data
synthesis problems beyond the reach of state-of-the-art solvers such
as Z3~\cite{enumerative-2022}, (3) the use of ACL2s to prove
refinement for optimized reactive systems, hardware and low-level
software~\cite{ manolios2019sks, manolios2015sks, manolios2015acl2,
  manolios2008refinement, Manolios2006AFF}, and (4) the development,
formalization and partial analysis of complete, executable formal
models for distributed gossip protocols, which were then used to find
flaws in applications (\eg, Ethereum) based on these
protocols~\cite{gossipsub-2022}. ACL2s supports the use of external
solvers and includes a systems programming capability that allows us
to use the core theorem prover as a key component in
formal-methods-enabled projects relating to gamified verification,
education, proof checking, interfacing with external theorem provers
and security~\cite{walter-acl2-systems-programming}.


\section{Conclusion and Future work}
\label{sec:conclusion}
We presented a Dijkstra-inspired calculational proof format with an
associated proof checker in ACL2s that has been used at Northeastern
University to teach over a thousand freshman-level undergraduate
students how to reason about computation.  For future work, we plan to
develop a ``professional'' version of the proof checker that is
interactive, automatically generates proof obligations (\eg, for
inductive proofs), proves easy cases automatically, and intelligently
generates aspects of proofs and hints.  We believe that such
functionality can be of immense use for proof engineers writing ACL2s
proofs.



\newpage
\bibliography{refs}

\end{document}